\documentclass[aip, amsmath,amssymb,reprint,nofootinbib]{revtex4-1}
\usepackage{graphicx,color,upgreek}
\usepackage{mathtools}
\usepackage{amsmath}
\usepackage{layouts}
\usepackage{float}
\usepackage{hyperref}
\hypersetup{
    colorlinks=true,
    linkcolor=blue,
    filecolor=magenta,      
    urlcolor=cyan,
    pdftitle={Overleaf Example},
    pdfpagemode=FullScreen,
    }
\usepackage{makecell}
\usepackage[abs]{overpic}
\usepackage{helvet}
\usepackage{graphicx,color,natbib}
\usepackage{placeins}
\usepackage[dvipsnames]{xcolor}
\usepackage{array, booktabs, makecell}
\usepackage{siunitx}
\usepackage{longtable}
\usepackage{tabularx}
\usepackage{booktabs}
\let\oldAA\AA
\renewcommand{\AA}{\text{\normalfont\oldAA}}
\usepackage{textcomp}
\usepackage{color}
\usepackage{amssymb,amsmath} 
\usepackage{siunitx}

\definecolor{cream}{rgb}{0.8706,0.8510,,0.7882}
\definecolor{strawberry}{rgb}{1.0,0.0,0.5}
\definecolor{midnight}{rgb}{0.00392,0.09804,0.57647}
\definecolor{blueberry}{rgb}{0.01569, 0.2, 1.0}
\definecolor{asparagus}{rgb}{0.4,0.4,0}
\definecolor{teal}{rgb}{0.0,0.5664,0.5742}
\definecolor{maraschino}{rgb}{1.0,0.0,0.0}
\definecolor{seagreen}{rgb}{0.15,0.48,0.27}
\definecolor{mygreen}{rgb}{0.0,0.55,0.3}
\definecolor{aqua}{rgb}{0, 0.5882,1.0}
\definecolor{tab}{rgb}{0.639215686,0.756862745,0.678431373}

\newcommand{\paddyspeaks}[1]{{\color{blueberry} #1}}

\newcommand{\rlj}[1]{{\color{mygreen}#1}}

\newcommand{\pecl}{\operatorname{\mathit{P\kern-.08em e}}}

\newcommand{\tausim}{\tau_{\rm sim}}

\renewcommand{\vec}[1]{\mathbf{#1}}

\begin{document}

\title{A model colloidal gel former for the measurement of inter-particle forces}

\noindent\large{

\author{Rui Cheng}
\affiliation{Gulliver UMR CNRS 7083, ESPCI Paris, Universit\' e PSL, 75005 Paris, France.}
\email{rui.cheng@espci.psl.eu}
\affiliation{H. H. Wills Physics Laboratory, University of Bristol, Bristol BS8 1TL, UK.}
\affiliation{Bristol Centre for Functional Nanomaterials, University of Bristol, Tyndall Avenue, Bristol, BS8 1TL, UK. }

\author{Kristian Thijssen}
\affiliation{Niels Bohr Institute, University of Copenhagen, Blegdamsvej 17, Copenhagen 2100, Denmark.}

\author{Abraham Mauleon Amieva}
\affiliation{H. H. Wills Physics Laboratory, University of Bristol, Bristol BS8 1TL, UK.}
\affiliation{Bristol Centre for Functional Nanomaterials, University of Bristol, Tyndall Avenue, Bristol, BS8 1TL, UK.}

\author{Tanniemola B. Liverpool}
\affiliation{School of Mathematics, University of Bristol, Bristol BS8 1UG, UK.}

\author{Malcolm A. Faers}
\affiliation{Formulation Design, Bayer AG, 40789 Monheim am Rhein, Germany.}

\author{Robert L. Jack}
\affiliation{Yusuf Hamied Department of Chemistry, University of Cambridge, Lensfield Road, Cambridge CB2 1EW, UK.}
\affiliation{DAMTP, Centre for Mathematical Sciences, University of Cambridge, Wilberforce Road, Cambridge CB3 0WA, UK.}

\author{C. Patrick Royall}
\affiliation{Gulliver UMR CNRS 7083, ESPCI Paris, Universit\' e PSL, 75005 Paris, France.}
\affiliation{Department of Physics, University of Warwick, Gibbet Hill Road, Coventry, CV4 7AL, UK.
}
\email{paddy.royall@espci.psl.eu}

\begin{abstract}
Tracking of particle coordinates in colloidal model systems is an established technique. Recently, an emulsion system has been introduced where not only the coordinates, but also contacts and compressive forces between the droplets can be determined [Dong \emph{et al.}, \emph{J. Chem. Phys.} \textbf{156} 214907 (2022)]. Here we develop this system to control both the density mismatch and refractive index contrast between the emulsion droplets and the solvent. The former is important in minimizing sedimentation due to gravity and the latter is necessary for 3d imaging with confocal microscopy. By mapping our system carefully to computer simulation, we quantify interdroplet forces and packing geometries in 
depletion gels and probe the relationship between local structure and compressive stress. The simulations also provide ensemble measurements of the attractive part of the force between droplets and related quantities such as the stress tensor. 
Finally, we implement an improved measure of local stress in our experiments, which can be directly compared with the corresponding quantity in simulations.  
We find that locally rigid minimum energy clusters are associated with regions of higher stress in the gel network.
\end{abstract}

\maketitle

\section{Introduction}
\label{sectionIntroduction}

Colloidal model systems provide great insight into the basic properties of a wide range of materials~\cite{poon2002,evans2019,royall2024}.
A particular attraction is real space imaging, where the coordinates of the constituent particles are tracked, providing an unprecedented level of information at the microscopic level~\cite{vanblaaderen1995,ivlev}. 
Such particle-resolved studies are ideal to study amorphous materials such as colloidal glasses and supercooled liquids~\cite{vanblaaderen1995,weeks2000,hallett2018} and gels~\cite{verhaegh1997,dinsmore2002,campbell2005,dibble2006,lu2008,royall2008,hsiao2012,royall2021,waheibi2024}. 
Among the requirements for particle-resolved studies in 3d is that the colloidal particles are refractive index matched to the solvent and that they do not sediment, which often means that their (mass) density should be matched to the solvent also~\cite{ivlev,royall2021}.

Colloidal gels {are important materials in their own right} that are used in everyday products including cosmetics, foods, pharmaceuticals, pigment printing inks, wound sealing, detergent formulations and cement 
~\cite{drury2003,rose2014,liang2001,ubbink2012,ioannidou2016,zaccarelli2007,ramos2005} and are also encountered in biological settings~\cite{darras2022}. Since they are far-from-equilibrium and dynamically arrested, gels evolve slowly in their energy landscape~\cite{zaccarelli2007}, and a complete understanding of these materials remains elusive. However, there has been important recent progress with gels, including their creep behaviour~\cite{leocmach2014,lockwood2023},  mechanism for rigidity~\cite{fenton2023,nabizadeh2023}, hierarchical structure~\cite{bantawa2023,tsurusawa2023} and stress relaxation under load~\cite{song2022}.

Gels are yield stress materials~\cite{bonn2017}, and exhibit dynamical features which precede yielding~\cite{aime2018}. One important class of colloidal gels consists of colloid-polymer mixtures~\cite{poon2002}. Here gelation takes place by spinodal decomposition in which the colloid-rich phase undergoes dynamical arrest, leading to the emergence of a solid gel~\cite{lu2008,zaccarelli2007}, and which also results in gel strands with thicknesses of several particle diameters~\cite{teece2011,zhang2013,verhaegh1997,huh2007,testard2011,zia2014}; this stands in contrast to gels with directionally dependent interactions which can form equilibrium or weakly aging gels, in which strand thicknesses are often a single particle diameter~\cite{zaccarelli2007,rice2012,song2022,bantawa2023}.

Despite the developments noted above, yielding and failure of gels is a key area in which further insight is called for~\cite{nicolas2018,bonn2017}. For example, failure may be sudden, after a long period where the gel appears to be unchanging
~\cite{starrs2002,bartlett2012,teece2011,secchi2014,bonn2017,cheng2026}. 
Although mechanical properties like yield stress and forces can be measured using rheology~\cite{bonn2017}, such bulk testing does not have access to the information of the distribution and propagation of stresses within the gel network. In this regard, computer simulations provide complete information \cite{nicolas2018,zaccarelli2007,bouzid2017,vandoorn2018,johnson2018,thijssen2023,bhaumik2025sm,bhaumik2025pre}.
However, simulating systems that are comparable in lengthscale and timescale to experimental samples is challenging. Like simulations, specially designed colloidal model systems can provide direct imaging into localized plastic events that underlie yielding, yet on experimental time- and length-scales~\cite{schall2007,besseling2007,cheng2026}. Stresses within the network are expected to play a pivotal role before and during yielding particularly in the context of localized irreversible rearrangements \cite{bartlett2012, aime2018, dong2022, cipeletti2020}.

Beyond tracking the coordinates of the particles, in the context of gel yielding, one can imagine obtaining the \emph{interparticle forces}. In principle this should be possible by measuring the coordinates, as well-characterised effective interaction potentials are known for many colloidal systems~\cite{likos2001}. However, although these can now be obtained more accurately than in the past~\cite{kawafi2025,bierbaum2017}, some measurement error remains, often at around 5\% of the particle diameter~\cite{kuerten2025}. In the case that the interaction potential  varies  slowly on this lengthscale, it is then possible to estimate the force based on measurement of the coordinates ~\cite{assoud2009}. However more generally and for the colloid-polymer mixtures of interest here, tracking errors impact this kind of approach very significantly, so that another method is called for~\cite{royall2007jcp}.

For granular materials (with particles up to cm in diameter), forces between particles have been measured~\cite{majmudar2005,brujic2007}. Of particular relevance here is a system of emulsion droplets of poly-dimethyl siloxane (PDMS) in which the \emph{solvato-chromic} dye nile red exhibits different fluorescent properties at contacts between droplets and in the interior of the droplet. Recently, some of us have developed an analogue of this system on the colloidal lengthscale. Leveraging STimulated Emission via Depletion (STED) super-resolution imaging, interparticle contacts and compressive forces between droplets were obtained~\cite{dong2022}.

Here, we develop this approach in three ways. Firstly, 
in the earlier work~\cite{dong2022}, the emulsion droplets were not density matched to the solvent, and so the duration of the experiments was limited. Here we show the addition of further components, in the form of a dense oil pentafluoro toluene (PFT) enables us to control the density mismatch between the emulsion droplets and the solvent. We simultaneously ensure the stability of the emulsion and optimize refractive index matching of the droplets and the solvent by changing the solvent composition. Secondly, we match carefully our experimental system with Brownian dynamics computer simulations. This is achieved by matching both the range and the depth of the interaction between the colloids~\cite{royall2007jcp,richard2018,razali2017,bhaumik2025sm}. This enables us to infer the distributions of the tensile component of the forces, as these are not measured directly in our experiments. Finally, we implement an improved measure of local stress in our experiments, which can be directly compared with the corresponding quantity in simulations,  which can accurately reflect the interparticle interactions~\cite{irving1950,thijssen2023}.

This paper is organised as follows.In Sec.~\ref{sectionMethods}, we describe our methodology. This is followed by the presentation of our results in Sec.~\ref{sectionResults}. We conclude in Sec.~\ref{sectionConclusion}.

\section{Methods}
\label{sectionMethods}

Our Methods section is organized as follows. Sections~\ref{sectionSynthesis}--\ref{sectionContacts} describe the experimental  model system. In particular we describe the synthesis of the PDMS emulsion in Sec.~\ref{sectionSynthesis}. In Sec.~\ref{sectionRefractive} we describe our approach to optimize refractive index matching between the emulsion droplets and the solvent. Our method to density match the emulsion is detailed in Sec.~\ref{sectionDensity}. The method we use to identify contacts in our images is briefly described in Sec.~\ref{sectionContacts}. In Sec.~\ref{sectionMapping} we describe mapping our experiments to the one-component Asakura-Oosawa model and further how this model is approximated by the (continuous) Morse potential that is appropriate for our computer simulations. The simulation methodology we outline in Sec.~\ref{sectionComputer}. The local stress analysis that we employ is described in Sec.~\ref{sectionStress}.

\subsection{Synthesis of PDMS emulsions}
\label{sectionSynthesis}

The PDMS emulsion droplets were synthesised through the hydrolysis and condensation of dimethyldiethoxysilane (DMDES) monomers, following the procedure of Elbers et al.~\cite{elbers2015}. The monomers constituting the oil phase were prehydrolysed with distilled water at a volume ratio of 2:5. The mixture underwent vortexing for 6 minutes at 1300 rpm and was gently agitated for 24 hours using a roller bank at 26 rpm. The prehydrolysed mixture was introduced into an ammonia solution, formulated with proportions of 6.8\% oil monomers and 22.7\% ammonia. The resulting dispersion, total volume 300 ml, was dispersed using an IKA Ultra-Turrax homogeniser equipped with a S25N 10G dispersing element at 10000 rpm for 5 minutes.

We allowed the droplets to grow in a standing bottle, and monitored their sizes with a Leica DMI3000B optical microscope. Typically, the diameter of droplets reached 3 $\mu$m within a day. We carried out dialysis on the
remaining dispersion to eliminate ammonia. To stabilise the droplets, we added 1 mM SDBS (sodium dodecylbenzene sulfonate) into the dialysis solution. The solution was replaced twice daily until reaching a stabilized pH of 7. Nile red dye was directly added to this emulsion product and left on a roller at 26 rpm for a day to dissolve. This dilute emulsion stock was left to concentrate under gravity for about 2 weeks.

To reduce polydispersity, we used a fractionation method that exploits the droplet-size dependence of the phase boundary in a depletion-induced phase transition, as described by Bibette \emph{et al.}~\cite{bibette1990, bibette1991}.
As the droplet diameter increases, the phase transition occurs for lower surfactant concentration.  Sodium Dodecylbenzene Sulfonate (SDBS) Surfactant was added gradually to cause flocculation in the order of droplet size . Large droplets were found to be trapped in the floc and then cream to the top. The creamed region was removed. More monodisperse samples could be obtained after a few repeats by separating big droplets (in the solid phase) from smaller ones (in the liquid phase). Additional surfactants can also stabilise the emulsion from coalescence, and increase the dye concentration at the droplet contacts. We added 1 mM SDBS in each round and observed the separation due to creaming using a Leica DMI3000B optical microscope. The process was stopped when no large droplets were found in the creamed phase, typically after five repeats.

To characterize the size and polydispersity of the emulsion droplets, we mixed an equal mass of emulsion concentrated stock and glycerol to generate a (approximately) refractive index matched system for confocal imaging. We assume the top dense layer reaches random close packing (taken as $\phi_\mathrm{rcp}\approx0.64$), allowing us to maintain the emulsion volume fraction at $\phi = 0.32$. We loaded approximately 10 \textmu l into a sample cell. The cell is composed of a glass slide and a cover glass, four edges are sealed by UV glue (Norland optical adhesive NOA 61). The coordinates of the droplets were determined using the Colloids package~\cite{leocmach2013}. To optimize the imaging, we set the fluorescence excitation and imaging parameters as specified in the Appendix.

The mean diameter of the droplets $d$ was determined to be 2.7 \textmu m with a calculated polydispersity of 10\% (obtained by dividing the standard deviation by the mean). These droplets are suitable for tracking, thus offering significant potential for particle-resolved studies. The continuous phase viscosity $\eta$ is 6 mPa$\cdot$ s. As our unit of time, we take the Brownian time $\tau_\mathrm{B} = d^\mathrm{3} \pi \eta \slash (8 k_\mathrm{B}T) \approx 11$ s for our parameters, where $k_\mathrm{B}T$ is the thermal energy. To screen electrostatic charges, we used 10 mM potassium chloride (KCl) corresponding to a Debye length of 6 nm, so that the particles can be regarded as close to hard spheres~\cite{royall2024,dong2022}.

The polymer used is xanthan, whose radius of gyration has been estimated to be  $R_g = 194$ nm in a similar system ~\cite{teece2011}. The polymer radius of gyration is a hard quantity to measure precisely and is sensitive to a number of factors such as solvent composition which we vary here. Here we follow previous work~\cite{dong2022,royall2018jcp} and treat $R_g$ is a (slightly) tunable parameter to obtain agreement with the fluid-gel phase boundary in the one-component Asakura-Oosawa model. While the AO model is itself an approximation, this approach is convenient for mapping to computer simulation (see Sec. ~\ref{sectionMapping}).

\subsection{Refractive Index Matching by Changing Solvent Composition}
\label{sectionRefractive}

Since the refractive index for the oil phase PDMS $n_\mathrm{PDMS} = 1.41$, we varied the solvent composition of water ($n_\mathrm{w} = 1.33$) and glycerol ($n_\mathrm{g} = 1.47$) to achieve an index close to that of oil droplets. The water mass ratio $r_\mathrm{w}$ in the solvent, $r_\mathrm{w} = m_\mathrm{w}/(m_\mathrm{w}+m_\mathrm{g})$, was prepared across a wide range, 0, 0.2, 0.4, 0.6, 0.8. Here $m_w$ and $m_g$ are the masses of water and glycol respectively. In order to assess the refractive index matching, we begin with visual observation. As shown in Fig.~\ref{figRImatching} (a), mass ratio $r_\mathrm{w}$ of 0.4 and 0.6 exhibited a clearer appearance. In order to achieve more precise matching, we focus on solvent compositions $r_\mathrm{w}$ between 0.4 - 0.6 (Fig. \ref{figRImatching} (b)) using a smaller step size of 0.05 in $r_\mathrm{w}$.

\begin{figure*}
\centering
\includegraphics[width = 160mm]{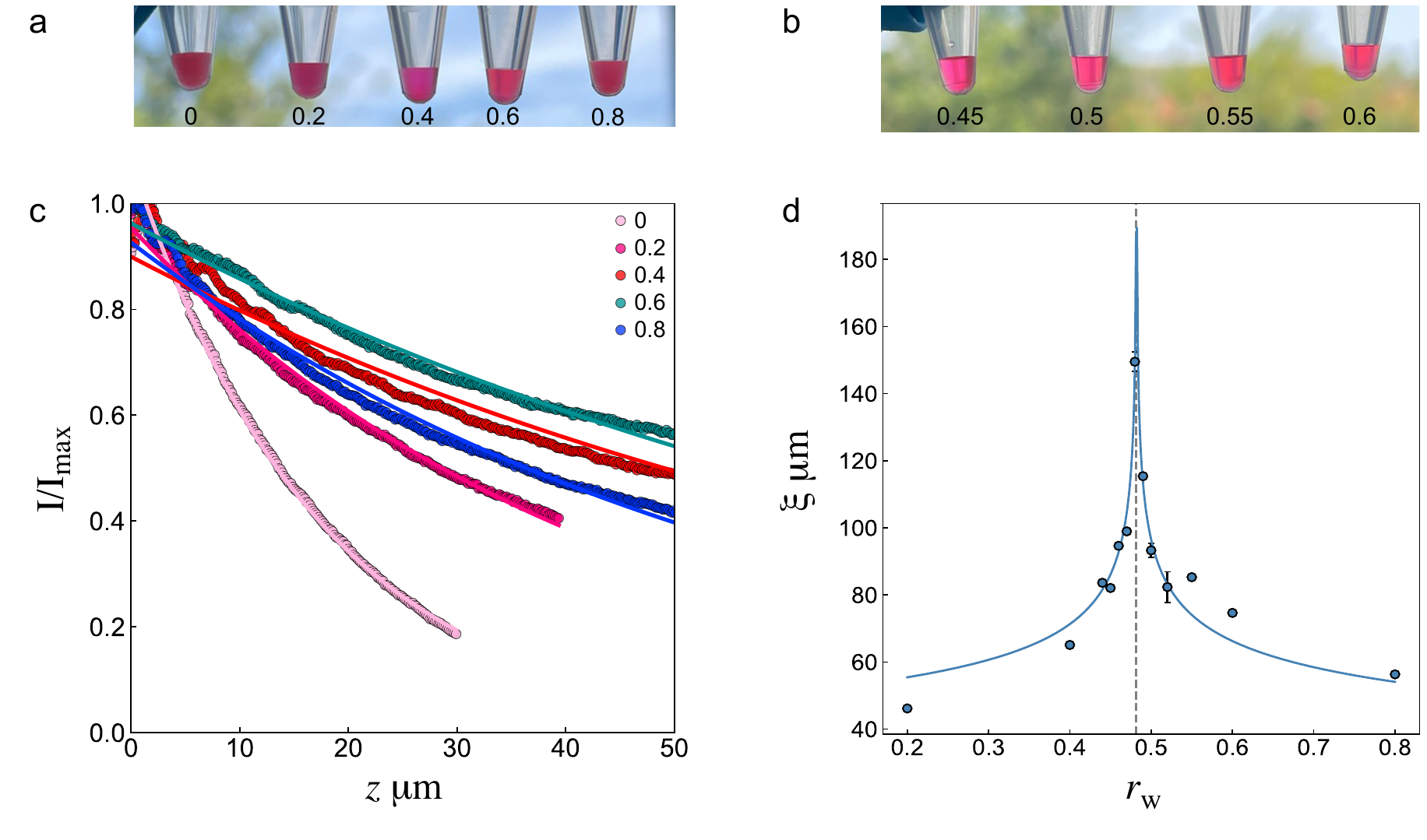}
\caption[Refractive index matching.]{(a-b) Samples at different water ratio $r_\mathrm{w}$. In all cases the emulsion volume fraction is around 0.3. (c)  Intensity at different $r_\mathrm{w}$, as a function of imaging depth $z$.
(d) The characteristic length at different solvent compositions. These are fitted with Eq.~\ref{eqRImatchingFit} from which we see that the optimal RI matching is at a water context of $r_w^\mathrm{opt}=0.482$.
}
\label{figRImatching}
\end{figure*}

To refine the RI matching, we measured the intensity profile along the axial direction ($z$) from confocal microscopy images. We loaded colloidal fluids ($\phi = 0.32$) into 50 $\mu$m glass cells. We allowed the samples to sit overnight, enabling the droplets to settle under gravity to reach equilibrium. The confocal laser was set at the same excitation wavelength (514 nm) and strength.

For each sample, we analysed five 3d images. In each image, we determined the intensity profile by plotting the intensity $I(z)$ averaged over $xy$ as a function of $z$, and normalised it by its maximum intensity $I_\mathrm{max}$. These curves were then fitted by:
\begin{equation}\label{eqRImatching}
I = A\exp(-z/\xi)
\end{equation}
where A is a constant, $\xi$ represents the characteristic length associated with the attenuation due to the RI mismatch between the droplets and the solvent. For each image, we obtained one $\xi$, Fig. \ref{figRImatching} (d) shows the mean value of $\xi$ at various compositions. The error bars are mostly smaller than the data points.

To evaluate the best RI matching we fit the length to 
\begin{equation}\label{eqRImatchingFit}
\xi = \frac{A}{|r_w-r_w^\mathrm{opt}|^\nu} + C
\end{equation}
\noindent
where $A$, $C$ and $\nu$ are constants and obtain that $r_w^\mathrm{opt}=0.482$ is the water content for optimal refractive index matching.

\subsection{Density Matching by Changing Droplet Composition}
\label{sectionDensity}

The relatively large colloids used in particle-resolved studies make density matching of the particles (or, in our case, droplets) essential for suppressing sedimentation. Importantly, we seek to retain the stability of the emulsions and also to retain conditions where the solvatochromic dye is effective and to avoid any leakage of the dye into the solvent and also to maintain the stability of the emulsion. Here we used 2,3,4,5,6-Pentafluorotoluene (PFT), ($\rho_\mathrm{PFT} = 1.44$). PFT furthermore is closely matched in refractive index (1.40) to PDMS (1.41), and insoluble in water but soluble in oil. The ratio $X_\mathrm{dense}$ represents the volume proportion of dense oil (PFT) to the total volume of the oil phase.

Besides the addition of PFT, the rest of the synthesis procedure remains the same as that of~\citet{dong2022}. We tested various $X_\mathrm{dense}$ to assess the density mismatch between colloids and the continuous index-matching solvent, as listed in Table \ref{tabGravitational}. A week after synthesis, we observed the left three samples in Table~\ref{tabGravitational} sediment, while the right two samples creamed. This indicates that the matching point lies between $X_\mathrm{dense}$ of 0.14 and 0.20. To quantify the density difference, we determine the gravitational length $l_{g}$  and subsequently calculate the density difference $\Delta\rho$ by fitting their sedimentation profiles \cite{royall2005}. The gravitational length is defined as $l_\mathrm{g} = k_\mathrm{B}T/\delta mg$, where $\delta m$ is the bouyant mass of a colloid. The length scale represents the height difference that occurs when the gravitational potential energy changes by the thermal energy $k_\mathrm{B}T$ \cite{perrin}. In dilute suspensions where colloids can be considered non-interacting (similar to an ideal gas), the gravitational length can be determined from the decay constant of the barometric height distribution, as described by an exponential decay of the volume fraction $\phi$ with height $z$ \cite{royall2005}: 
\begin{equation}\label{eqDensityMatching}
\phi(z) = \phi_{0}e^{-z/l_\mathrm{g}}
\end{equation}
where $\phi_{0}=\phi(z=0)$ is the volume fraction at the start of the profile.

To conduct this experiment, we diluted the concentrated emulsion stock to 10\% colloid volume fraction. Both the emulsion stock and the solvents contained 10 mM KCl to screen electrostatic charges. The samples were loaded into glass cells of height 30 \textmu m, and left to settle into an equilibrium density profile. Normally overnight, particles sediment to form a dense layered region at the bottom of the cell (or top, if they cream), an example image in the $xz$ direction is shown in Fig.\ref{figDensityMatching} (a). Above (or below) this region and in the dilute regime ($\phi\to 0$), the density decreases following a barometric law, this is the region where we are interested. For each sample, we imaged 50 3d confocal images. After tracking using the Colloids package \cite{leocmach2013}, we obtained coordinates of the particles and a measure of their radii.

The droplet volume fraction $\phi$ was obtained by binning the identified partial/whole sphere volume in the $z$ dimension. For particles at low enough volume fraction (here $\phi \leq 0.03$), the equilibrium density profile follows the barometric law. This is plotted 
in Fig. \ref{figDensityMatching} (b). The density decrease or increase with height corresponds to colloids sedimenting or creaming respectively. By fitting the slope for each sample we obtained the gravitational length from Eq.~\ref{eqDensityMatching}. The results are presented in Table~\ref{tabGravitational}.

\begin{table}
\begin{center}
\begin{tabular}{ c c c c c c c}
\hline
Fraction of dense oil  $X_\mathrm{dense}$ & 0.33 & 0.20 & 0.14 & 0.11 & 0.09 \\ 
\hline
Gravitational length  $l_{g}/d$ & 4.98 & 10.98 & 18.28 & -7.71 &  -5.03 \\ 
\hline
Density difference $\Delta\rho (g/cm^{3})$ & 0.05 & 0.02 & 0.01 & -0.05 & -0.03 \\ 
\hline
\end{tabular}
\caption[Gravitational length.] {Density matching for the PDMS-PFT emulsion. The gravitational length $l_\mathrm{g}$ is found by fitting curves from Fig. \ref{figDensityMatching} (b). The density difference $\Delta\rho$ is calculated from the gravitational length, it turned from positive (sediment) to negative (cream) with more light oil in. 
}
\label{tabGravitational}
\end{center}
\end{table}

Here we seek to suppress rather than to eliminate the effects of sedimentation as complete elimination is challenging for droplets of this size~\cite{royall2005}. Thus, we take the $X_\mathrm{dense} = 0.14$ sample, which exhibited a gravitational length of approximately 18$d$ for colloids with a mean diameter of 2.7 \textmu m. The density mismatch in this case is around  0.01 gcm$^\mathrm{-3}$ which is a marked improvement over previous work~\cite{dong2022}. Further details of fluorescent characterization are provided in the Appendix.

\begin{figure}[t!]
\centering
\includegraphics[width = \columnwidth]{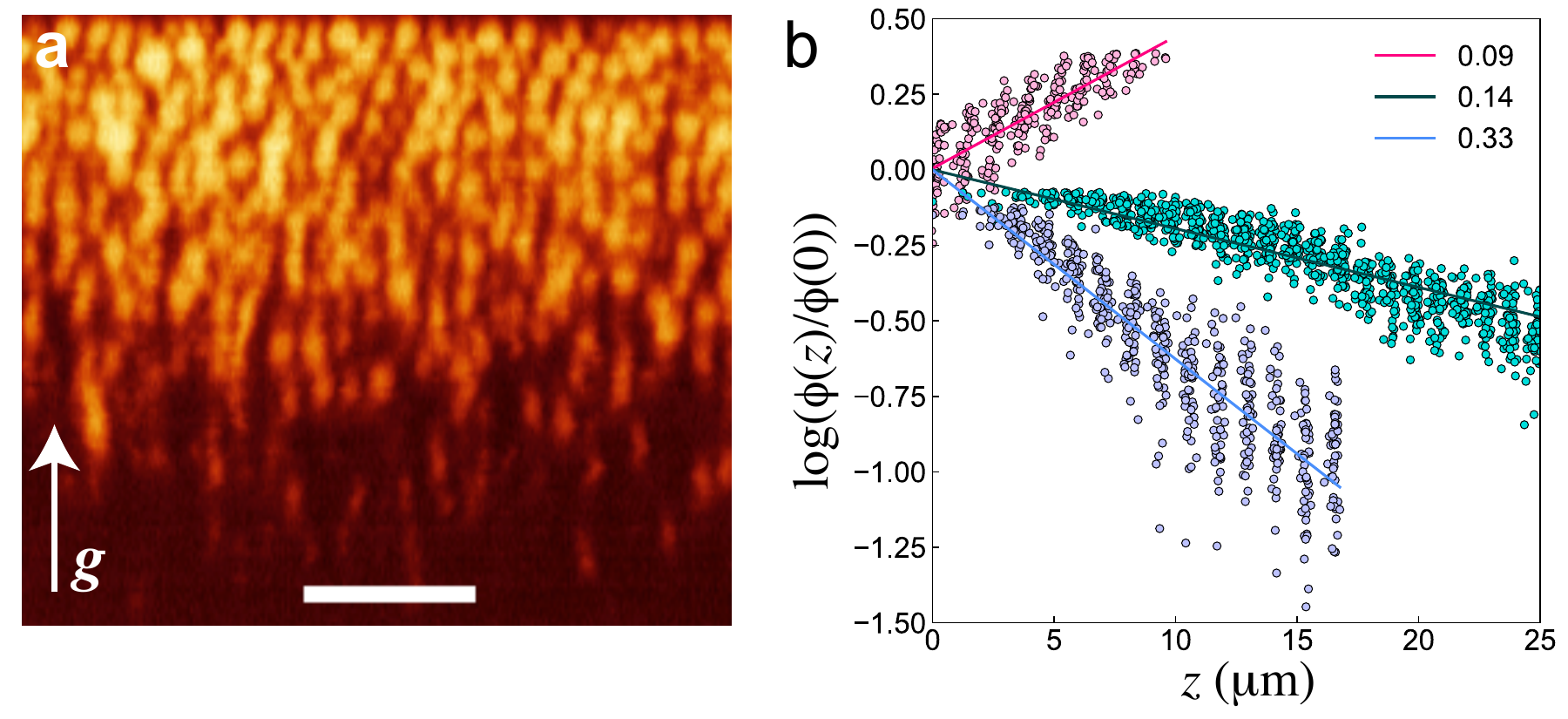}
\caption[Density matching.]{Density matching by adjusting the droplet composition.
(a) An example slice image in the $xz$ direction, 3D stacks were used to extract the density profiles. Here the droplets cream as indicated by the direction of $g$. Bar=10~$\mu$m.
(b) Log volume fraction at different oil ratios $X_\mathrm{dense}$. Data are scaled by the value at the start of the profile. Points are from tracking the time series experiments, lines are from fitting Eq. \ref{eqDensityMatching}.
}
\label{figDensityMatching}
\end{figure}

\subsection{Contacts from Images}
\label{sectionContacts}

Here we summarise the contact tracking method used, which is developed slightly from that in Ref.~\cite{dong2022}. We employ two Otsu filters: the first distinguishes the background from the foreground elements, and the second separates the droplet centre fluorescence from the real contacts.

We first computed the weighted middle points positioned between droplets, in principle on the line connecting the centers of droplets, and the distance to each center is proportional to the respective droplet radii. After this we generated a sphere-like binary structure from the center of middle points, and then dilated the structure. We iterate this process four times.

Each contact is assigned an index $j$ and is stored with its contact volume $v_j$ (determined by counting the number of pixels {associated with the contact}). The position of each contact is given by its geometrical centre. We then allocate these contacts to their closest pair of droplets based on their distances. Specifically, the vectors between contact $j$ and the two closest droplets are $\vec{R}_{ij}$ for  $i=1,2$.  The lengths of these vectors are denoted $R_{ij}$ for $i=1,2$: these can be identified with the radii of the two droplets. In turn we define the number of contacts $N_\mathrm{con}$ for each droplet.

\subsection{Mapping to Theory and Simulation} 
\label{sectionMapping}

We discuss here the effective polymer-induced attractions between the emulsion droplets, and how they are modelled in our simulations.  
We describe them using the 
one-component Asakura-Oosawa (AO) model~\cite{asakura1954,royall2008,poon2002,lekkerkerker}, which reads
\begin{widetext}
    \begin{equation}
\beta u_\mathrm{AO}(r)=
\begin{cases}
\infty & \mathrm{for~} r < d \\
\frac{4\pi R_g^3 z_p^r(1+q)^3}{3q^3}  
\left[ 1-\frac{3r}{2(1+q)d}+\frac{r^3}{2(1+q)^3d^3} \right] & \mathrm{for~} d \le r < d+2R_g \\
0 & \mathrm{for~} r \ge d+2R_g
\end{cases}
\label{eqAO}
\end{equation}
\end{widetext}
where $\beta=1/(k_BT)$ and the polymer-colloid size ratio $q=2R_g/d$.     For ideal polymers, the \emph{fugacity} $z_p^r$ is equal to their number density in a reservoir at the same chemical potential as the colloid-polymer mixture.

The accuracy of the AO model for this system has been verified in the equilibrium fluid regime for state points similar to those considered here~\cite{royall2007jcp}. To parameterise the AO potential we fix $z_p^r$ equal to the polymer concentration $c_p$ and we treat $R_g$ as an adjustable parameter, which we determine via the extended law of corresponding states~\cite{noro2000}.  This law states that the reduced second virial coefficient at the critical point for phase separation obeys
\begin{equation}
B_2^*=\frac{3}{\sigma_\mathrm{eff}^3} \int ^\infty _0 r^2 [1-\exp(-\beta u(r))]dr \approx -\frac{3}{2}
\label{eqb2star}
\end{equation}
where $\sigma_\mathrm{eff}$ is the effective hard-sphere diameter.  (We neglect colloid polydispersity for the purposes of these calculations.)
We estimate in Sec.~\ref{sectionPhase} below that the critical point is at $c_p^*\approx 0.03 \mathrm{g \, l}^{-1}$: combining this with from which Eqs.~(\ref{eqAO},\ref{eqb2star}) yields $q=0.182$. This value of $q$ is approximately 20\% larger than that estimated by~\citet{bartlett2012}. 
To rationalise this we note that accurate experimental determination of $R_g$ is challenging, and in any case we have changed the solvent composition with respect to Ref.~\onlinecite{teece2011}, which will impact $R_g$.  Moreover, we have assumed ideal AO behavior, which is itself an approximation~\cite{lekkerkerker}.

Obtaining an effective AO model in this way also relies on several other assumptions.  In particular, we estimate $c_p^*$ from the onset of gelation and not directly from the experimental critical point: this is largely justified because the onset of gelation is rather insensitive to volume fraction~\cite{royall2018jcp}.  For high polymer concentrations (comparable and beyond the overlap concentration), one also expects deviations from AO behaviour because the dominant lengthscale becomes the persistence length of the polymer rather than the radius of gyration of the polymer coil as assumed in the AO model.  However, accurate computer simulations have shown that these deviations are small for the state points considered here~\cite{louis2002jcp} (the maximum polymer concentration is approximately 2.8 times that of overlap).

For the purposes of computer simulation, we follow~\cite{royall2008, taffs2010, zia2014, landrum2016, razali2017, griffiths2017} and mimic the AO potential by an attractive Morse potential.  An accurate mapping between these potentials was made in Ref.~\onlinecite{taffs2010}. It is convenient in molecular dynamics simulations because the Morse potential is continuous, in contrast to the AO model which has an infinitely repulsive hard core.  It reads
\begin{equation}
    \beta u_\mathrm{Morse}(r)=\beta  \varepsilon\big[e^{-2\alpha(r-d_{ij})}-2e^{-\alpha(r-d_{ij})} +c_{\rm sh} \big]  
\label{eqU}
\end{equation}
for $r<r_c$, and $\beta u_\mathrm{Morse}=0$ otherwise,
where $d_{ij}$ is the average diameter of particles $i$ and $j$; the interaction strength is $\varepsilon$ and $\alpha$ sets the range; the cutoff parameter is $r_c=1.4 d_{ij}$, and we truncate so that $u_\mathrm{Morse}(r)=0$ for $r>r_c$; the constant shift $c_{\rm sh}$ is chosen so that $\beta u_\mathrm{Morse}$ is continuous at $r= r_c$.

To match the AO interaction with the Morse potential, we set the depths of their attractive wells to be equal: 
\begin{equation}
\beta\varepsilon=\frac{4\pi R_g^3 z_p^r}{3}.
\label{eqEpsilon}
\end{equation}
Using again the extended law of corresponding states in Eq.~\ref{eqb2star} so that the critical points of Morse and AO systems coincide, we obtain $\alpha\approx39.048{/d}$. 

The Morse interaction strength at the critical point is then $\varepsilon^* = 3.337k_{\rm B}T$.  This means that the experimental polymer concentrations of $c_p=[0.08, 0.1, 0.16, 0.2, 0.35]$ gl$^{-1}$ map to $\varepsilon=[8.9, 11.1, 17.6, 22.3, 39.0]k_{\rm B}T$. 
For certain parts of the data analysis, we also consider two state points below the gelation threshold,  $\varepsilon=[2.6, 3.3]k_{\rm B}T$.

\subsection{Computer Simulation}
\label{sectionComputer}

For simulations, we use a size-polydisperse system comprising seven particle sizes, with diameters equally spaced between $d_{ij}=0.841 d$ and $d_{ij}=1.159 d$ between particles $i$ and $j$ to match the experimental polydispersity of 10\%.  We fix an equal mass across all particle diameters. We use the mean particle diameter $d$ as the unit of length and define the time unit as $\tau_{\rm sim}=(m d^2/k_{\rm B}T)^{1/2}$, with $\tau_{\rm sim}=1$ in reduced units. The number of particles $N$ in a cubic box of volume $L^3$ is chosen to match the corresponding experimental volume fraction, which is calculated in the simulations as $\phi=\pi N{\bar{d^3}}/(6L^3)$, where we use the averaged $\bar{d^3}=\left\langle d_{ij}^3 \right\rangle$.

The colloidal dynamics are modeled using Langevin dynamics with a large friction coefficient to approximate the overdamped regime. The equation of motion for particle $i$, with position $\mathbf r_i$, velocity $\mathbf v_i$, and mass $m$ is
\begin{equation}
m\frac{d\mathbf{v}_i}{dt}=-\nabla_i V -\lambda \mathbf{v}_i  +\sqrt{2\lambda k_B T}\mathbf{\xi}_i
\label{eqLan}
\end{equation}
where $V= \sum_{1\leq i < j\leq N} u_\mathrm{Morse}(|\mathbf{r}_i-\mathbf{r}_j|)$ is the total potential energy and
$\lambda$ is the friction constant. The non-dimensionalized friction is $\lambda_0 = \lambda \tausim/m$.  We take $\lambda_0=10$ to mimic overdamped dynamics~\cite{bhaumik2025sm,razali2017}.
The integration time step is $\Delta t=0.001 \tausim$. To map to the timescales in the experiments, we note that here that the Brownian time is $\tau_B = 0.417\tausim$.

This simulation method is computationally efficient and has been shown to capture the essential features  of depletion gels~\cite{zia2014,landrum2016}, including quantitative comparisons with experiment~\cite{royall2008, razali2017,bhaumik2025sm}.  We note that our model neglects hydrodynamic interactions, which do affect some aspects of gel structure~\cite{royall2015,varga2015,degraaf2019}. All simulations are performed in LAMMPS~\cite{thompson2022}.

Simulations are initialized in random configurations.  Gels are formed by simulating for a time $t_{\rm w}$, during which spinodal decomposition occurs. Unless otherwise stated we simulate $t_{\rm w}=5\times 10^4\tausim$ ($\approx 1.2\times 10^5\tau_B$), at which point the gel has become arrested and no obvious further aging takes place over the time scales of our measurements; all results are averaged over several independent runs (typically 5), to ensure statistically robust conclusions.

\subsection{Stress Analysis}
\label{sectionStress}

Quantifying contact forces is a pivotal objective in this investigation. As noted in Sec.~\ref{sectionContacts}, each contact $j$ is linked to a pair of neighboring droplets according to a minimum distance rule. We use the Princen model to obtain a measure of the compressive force in each droplet, and then carry out the same analysis for simulations and experiments. Each contact $j$ involves two particles which we label as $i=1,2$; the radius of particle $i$ is $R_{ij}=|\vec{R}_{ij}|$, as defined in Sec.~\ref{sectionContacts}. We estimate the contact forces from the size of overlap (ie the contact regions) as
\cite{princen1983}:
\begin{equation}
\mathbf{F}_{ij} = \frac{\gamma}{\tilde R_j} A_j \frac{\mathbf{R}_{ij}}{R_{ij}}
\label{eqPrincen}
\end{equation}
where $\gamma=9.2 \text{mN/m}$\cite{jorjadze2013} is the interfacial tension of the emulsion droplets, $A_j$ is the deformation area due to the contact forces (assumed to scale with the contact volume $v^{2/3}$), and $\tilde R_j = \frac{2R_\mathrm{1j}R_\mathrm{2j}}{R_\mathrm{1j}+R_\mathrm{2j}}$ is a suitably weighted average radius. We later compare this (purely repulsive) effective force with a corresponding quantity extracted from simulations.

Our measurements of local stress use a volume-averaged representation of the Irving-Kirkwood (IK)stress~\cite{yang2012,smith2017,hardy1982}. In simulations,  we know $\vec{p}_i$ the momentum of particle $i$, and also $\vec{r}_{ij}$ for the vector connecting particles $i$ and $j$, and we know the total corresponding interparticle forces $\vec{f}_{ij}$.
Now consider a spatial region $\Omega$ whose volume is $|\Omega|$.  The components of the IK stress for that region are then \cite{smith2017}
\begin{equation}
\sigma^{\mu\nu}= \frac{1}{|\Omega|} \sum_{i=1}^N \bigg[ \frac{1}{m_i}p_i^\mu p_i^\nu\vartheta_{i,\Omega}+\frac{1}{2}\ \sum_{j(\neq i)} {r}_{ij}^\mu{f}_{ij}^\nu\varphi_{ij,\Omega} \bigg] ,
\label{eqIK}
\end{equation}
where $\vartheta_{i,\Omega}=1$ if particle $i$ is in $\Omega$ and zero otherwise; similarly $\varphi_{ij,\Omega}$ is the fraction of the straight line connecting particles $i,j$ that lies within $\Omega$.

Taking $\Omega$ to be the entire simulation box gives the total stress $\Sigma$, which can also be computed from the virial. For a local measurement of stress at point $\vec{r}$, we take $\Omega$ to be a small cube of side {$l_{\rm IK}$}, centered at $\vec r$.  The resulting stress is denoted by $\sigma(\vec{r})$. For local measurements, we take $l_{\rm IK} = 1.5 \bar d$, which is sufficiently small to allow a local measurement, but sufficiently large to avoid  uncertainties due to thermal fluctuations as demonstrated in previous work~\cite{thijssen2023}.

From this, we can assign to each particle a local stress, which is the value of $\sigma$ associated with the cube $\Omega$ that contains that particle. We find that the particle momenta do not contribute much to this local stress, which facilitates comparison to the experiments, where momentum is negligible. We report the trace of the stress tensor, denoted by $\sigma_{\mathrm{full}}=\mathrm{Tr}(\sigma)$, which includes all particle interactions.  This is in contrast to $\sigma_\mathrm{rep}$, which only includes repulsive interactions, see below.
We note that $\sigma_{\mathrm{full}}$ corresponds to three times the local (osmotic) pressure.

For comparison with experiment, we note that the Princen model for contacts [Eq.~\eqref{eqPrincen}] estimates repulsive interparticle forces, but is not sensitive to the attractive depletion forces.  To mimic this, we define the repulsive force between particles $ij$ as $\vec{f}_{ij}^\mathrm{rep} = \vec{f}_{ij} \Theta(d_{ij}-|\vec{r}_{ij}|)$ where the Heaviside function $\Theta$ restricts to particles that are closer than $d_{ij}$ so that this force is indeed repulsive. The repulsive part of the stress tensor is obtained by replacing $f_{ij}$ in Eq.~\eqref{eqIK} with $f_{ij}^\mathrm{rep}$ and we define $\sigma_\mathrm{rep}$ as the trace of this object, which is analogous to $\sigma_{\mathrm{full}}$.

An experimentally accessible measure of stress is found by using this same stress tensor, but summing over the contacts rather than the particles:
\begin{equation}
\sigma^{\mu\nu}_\mathrm{exp}= \frac{1}{|\Omega|} \left[\frac{1}{2}\sum_{j} \sum_{i=1}^2 {R}_{ij}^\mu{F}_{ij}^\nu\varphi_{ij,\Omega} \right] \; ,
\label{eqIKexp}
\end{equation}
where we again have boxes of about 1.5 the particle size and ascribe the trace of the experimental stress to every particle in that box. The index $j$ in Eq.~\eqref{eqIKexp} labels the contacts and the sum runs over all of them; the sum over $i$ accounts for the two particles involved with each contact.

The experimental measurement is not instantaneous, due to the image acquisition time (typically $\sim 5$ s for a droplet). We therefore average the simulation data over 5 $\tau_B$ to facilitate the comparison between experiment and simulation. Apart from improving the statistical accuracy, the choice of averaging procedure did not appear to affect the results when averaging was performed over $0.1 \tau_B$.

\section{Results}
\label{sectionResults}

Our results section is organized as follows. In Secs.~\ref{sectionPhase} and ~\ref{sectionStructural}, we characterize our system focusing on the phase behavior and structure. In the latter, we compare experiment and simulation, where the phase behavior of the simulations matches that of the experiments through our mapping procedure (Sec.~\ref{sectionMapping}).  In Sec.~\ref{sectionStressDistribution}, we analyze the stress at the local level in both experiments and computer simulations. In the final 
results section, we probe the relationship between locally rigid clusters and stress in Sec.~\ref{sectionLocal}.

\subsection{Experimental Phase Diagram}
\label{sectionPhase}

\begin{figure}[t!]
\centering
\includegraphics[width = \columnwidth]{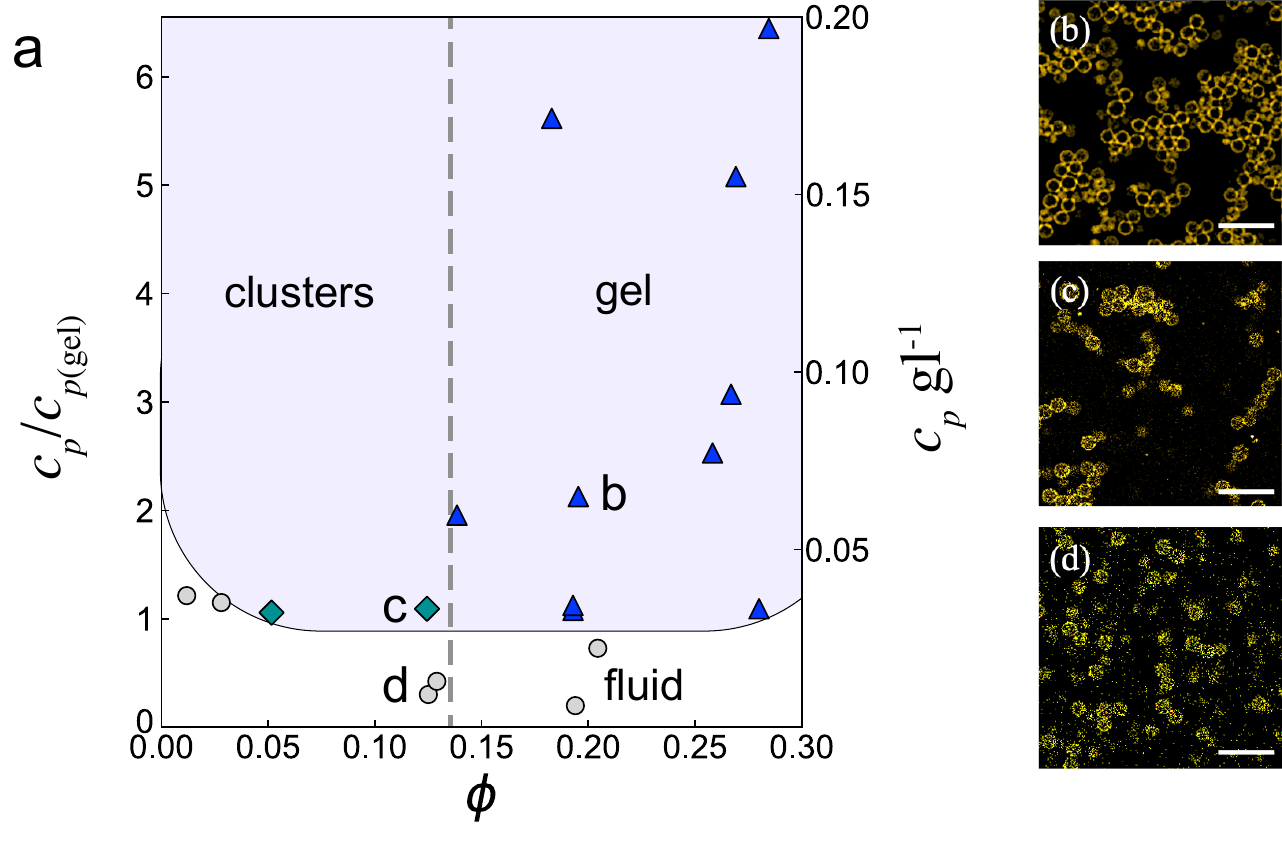}
\caption{\textit{Phase diagram of the 
emulsion volume fraction-polymer concentration system.}
(a) Phase diagram in the colloid volume fraction - polymer concentration plane. Gel, (non-percolating) clusters and fluid (F) regions are indicated. The data are plotted both as the absolute polymer concentration $c_p$ (right) and that scaled by the polymer concentration at gelation $c_p/c_{p\mathrm{(gel)}}$. Gels are denoted blue triangles,clusters as green diamonds and fluids as gray circles.
(b-d) are confcoal images of the states observed. 
(b) Gel.
(c) Clusters.
(d) Fluid. The locations of (b-d) are indicated approximately in the phase diagram (a). 
}
\label{figPhase}
\end{figure}

To position the samples we study in the context of the gelation phase boundary, and to facilitate comparison with
simulation, we determined a phase diagram by varying the volume fraction of emulsion droplets $\phi$
and polymer concentration $c_p$ as illustrated in Fig.~\ref{figPhase}. STED confocal images from the contact channel are presented on the right side, illustrating different states according to various positions on the phase diagram. (a) low $\phi$ and $c_p$ result in colloidal fluids, (b) low $c_p$ with relatively higher $\phi$ leading clusters which appear not to percolate in the imaging volume, (c) high $\phi$ and $c_p$ lead to a gel network with clusters percolating the entire image.

We have identified the polymer concentration corresponding to gelation $c_{p\mathrm{(gel)}}$ (the bottom of the shaded region) to be 0.03 gl$^\mathrm{-1}$.
The volume fraction at which percolation in colloidal gels occurs is a function of time and system size~\cite{manley2005glasslike,griffiths2017}. For our parameters, 
we find that for an emulsion volume fraction $\phi \gtrsim 0.13$, we obtain a network that percolates within the imaging volume.

\subsection{Structural Characterization (Experiment)}
\label{sectionStructural}

The radial distribution functions $g(r)$ are depicted in Figure \ref{figG}. The first peaks rise subtly with increasing polymer concentration~\cite{royall2007jcp}. The first minimum (at a distance of $1.3d = 3.7$ \textmu m), served as a distance threshold for identifying neighboring particles \rlj{(see below)}.  Of the five state points shown in Fig.~\ref{figG}, we see a change in structure between $c_p/c_{p\mathrm{(gel)}}$=3 and 5 as a second peak emerges for higher $c_p/c_{p\mathrm{(gel)}}$.
The presence of a second peak indicates densely-packed particle arrangements in the system. For the purposes of the contact analysis, we found there to be too much residual Brownian motion for $c_p/c_{p\mathrm{(gel)}} \lesssim3$ due to the low packing of these droplets. Hence, we focus our attention on gels at three polymer concentrations: 0.16, 0.20, 0.35 gl $^\mathrm{-1}$, corresponding to 5.3, 6.7, 11.7 times the gel boundary $c_{p\mathrm(gel)}$.  We refer to the other state points as ``weak gels''.

\begin{figure*}[t!]
\centering
\includegraphics[width = 150 mm]{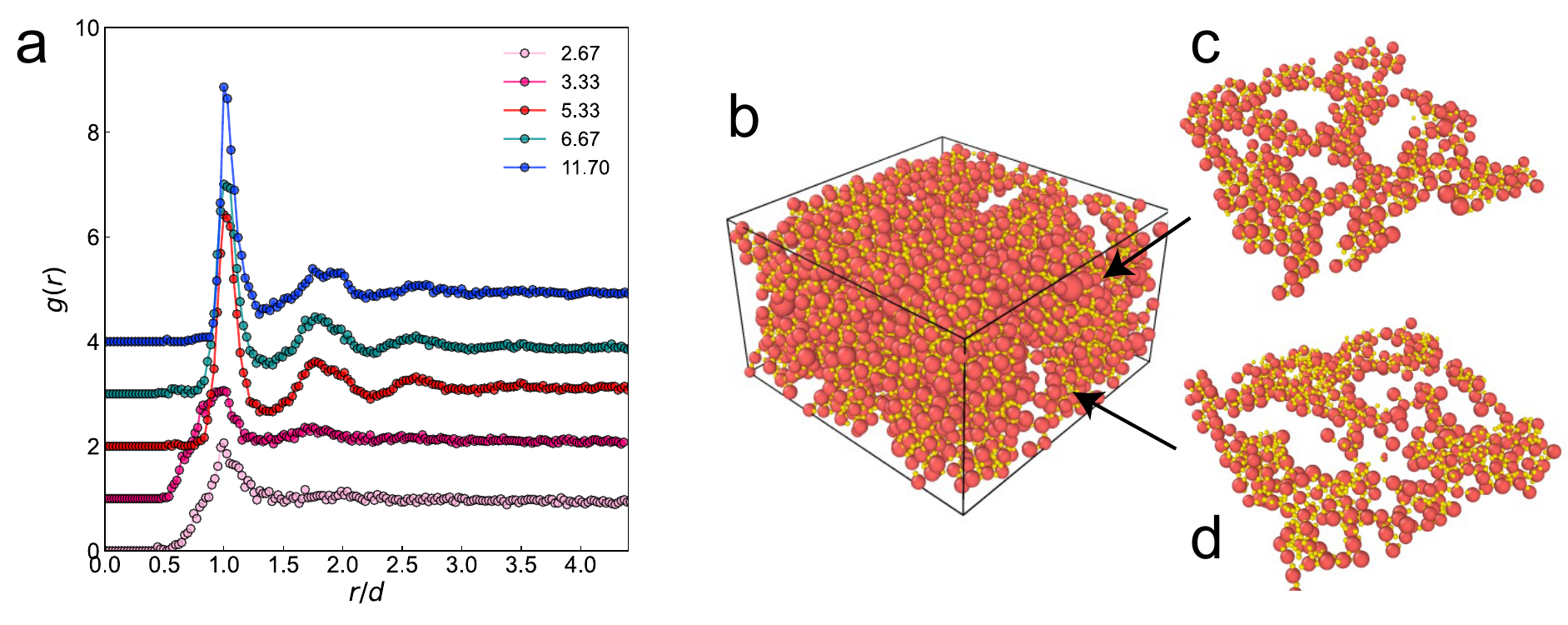}
\caption{\textit{Radial distribution functions at different gel strengths.}
(a) Radial distribution functions $g(r)$ at various polymer concentrations $c_p/c_{p\mathrm{(gel)}}$. (b) Reconstruction of droplets (salmon) and contacts (yellow) from the tracking results. (c) and (d) are the same data as (b) but only show slices of one-diameter-thickness from the top and the bottom for clarity. The renderings are from the experimental dataset of $c_p/c_{p\mathrm(gel)} = 6.7.$}
\label{figG}
\end{figure*}

To verify the tracking results, we rendered the droplets as colored particles and contacts as smaller spheres. As can be seen from Figure \ref{figG}(b-d), the yellow contacts are located between droplets in a reasonable manner.

{Example images of gels are shown in Fig.~\ref{figPerky} for a range of polymer concentrations. These clearly show the contacts between the particles and the ``arms'' of the gel that span these images. These images also show characteristically different morphology exhibited by the gel ``arms''.

\begin{figure}[t!]
\centering
\includegraphics[width = 75mm]{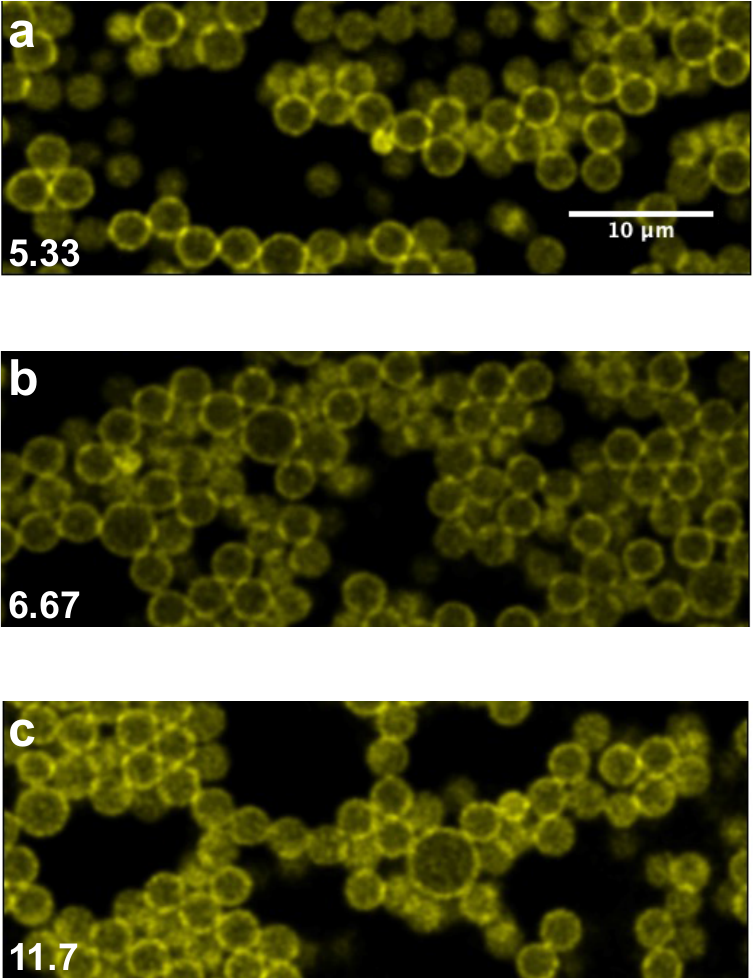}
\caption{\textit{Confocal images of percolating chains.} %
Snapshots of different physical appearance of gel arms (a) Single-string-like gel arms. (b) More compact and thicker gel arms. (c) Larger droplets with more neighbors. Scale bars indicate 10 \textmu m.
}
\label{figPerky}
\end{figure}

\begin{figure*}
\centering
\includegraphics[width = 140mm]{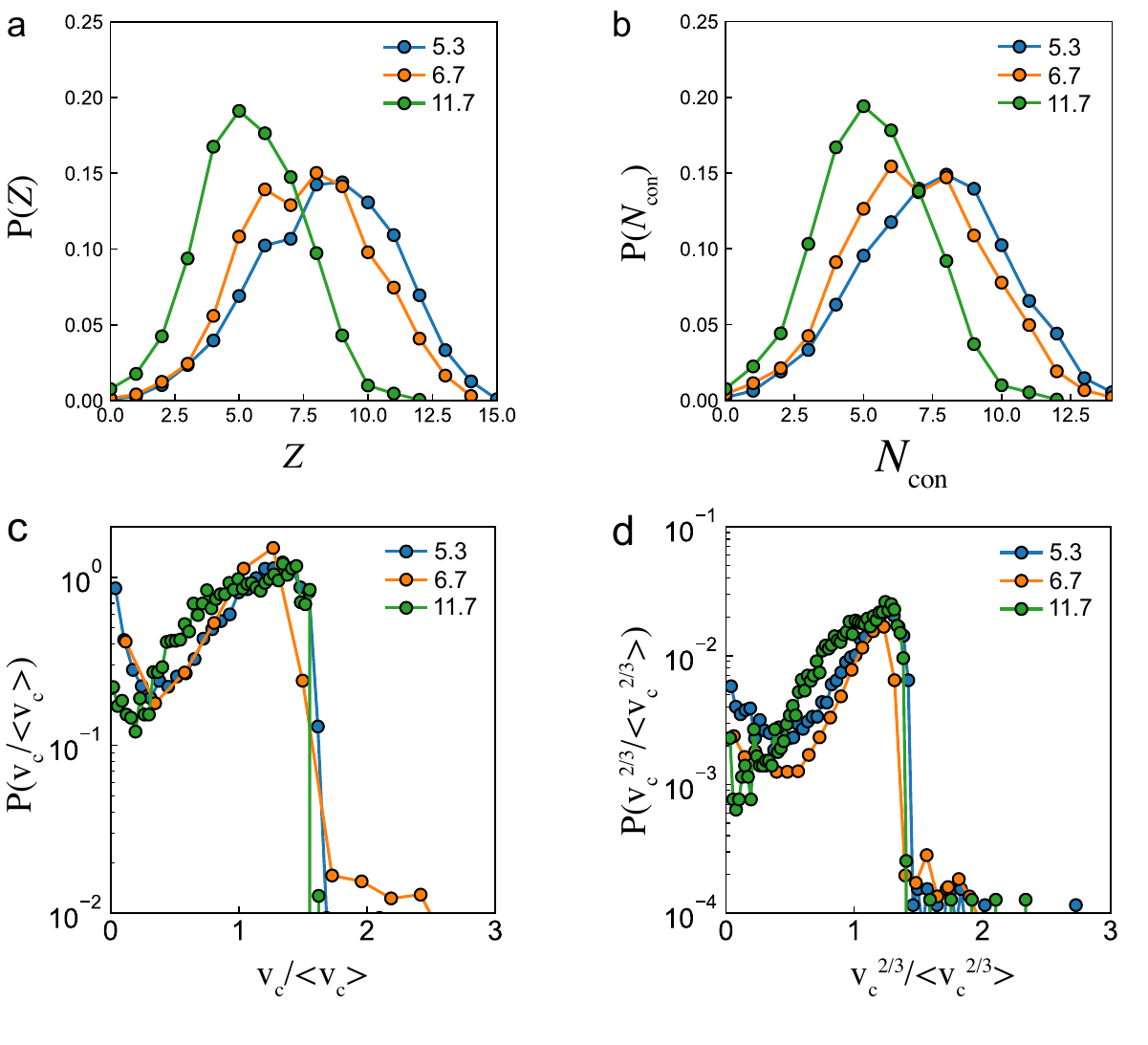}
\caption{\textit{Neighbors and contacts in experimental data.}
Legends indicate polymer concentrations $c_p/c_{p\mathrm{(gel)}}$.
(a) Distributions of number of neighbors $Z$. 
(b) Distribution of the number of contacts per particle $N_\mathrm{con}$.
(c) Distribution of contact volumes $v$ scaled by the mean.
(d) Distribution of contact ``areas'' $v^{2/3}$ scaled by the mean.}
\label{figNeighboursExp}
\end{figure*}

\begin{figure*}[t!]
\centering
\includegraphics[width = 130mm]{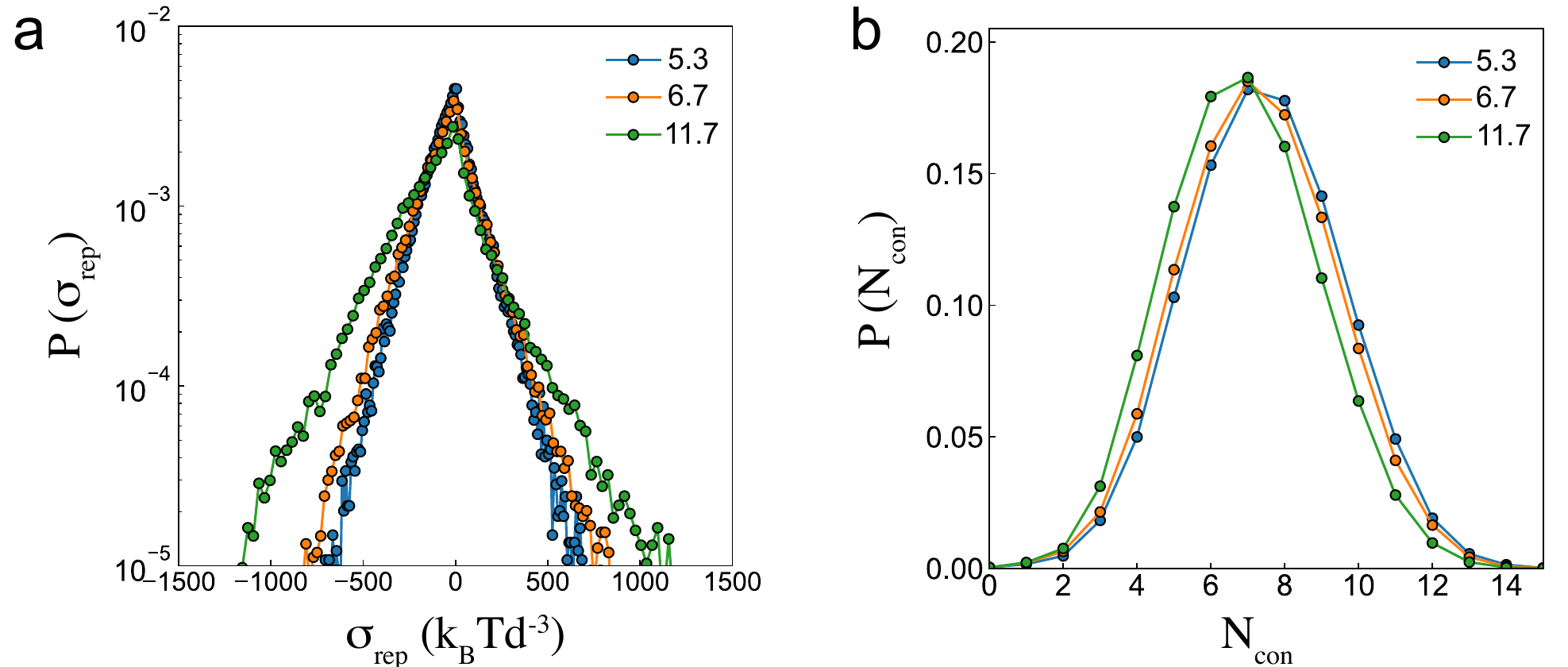}
\caption{
\textit{Distributions of local stress and number of contacts per particle in simulation data.} Legend shows the corresponding $c_p/c_{p\mathrm{(gel)}}$ values that were mapped to morse potential depths $\varepsilon=[17.6, 22.3,39.0]k_{\rm B}T$. (a) The local stress of particles inferred from all forces.
(b) Number of neighboring particles that are closer than 1.3 $\sigma$.
In our simulations, neighbours and contacts are identical by construction.
} 
\label{figNeighboursSim}
\end{figure*}

We then investigate the number of contacts $N_\mathrm{con}$ and neighbors $Z$ for each particle.  The contacts are defined as in Sec.~\ref{sectionContacts}, and we extract the number $N_\mathrm{con}$ of contacts in which each particle participates.  Similarly, two particles are neighbours if their separation is less than $1.3d$ and we extract the number of neighbours $Z$ of each particle. This analysis allows us to understand the connectivity within the gel system. Previous studies in a have demonstrated a correlation between particle mobility and $Z$, especially in weakly quenched gels~\cite{dibble2008,ohtsuka2008}.

In Fig.~\ref{figNeighboursExp}(a), the probability distributions of the number of neighbors $Z$ is presented. For the gel with $c_p/c_{p\mathrm{(gel)}}=5.3$ 
a large fraction (around 20\%) of droplets have five neighbors. For gels with a deeper quench, i.e. a higher polymer concentration ($c_p/c_{p\mathrm{(gel)}}=6.7$ and $11.7$), more neighbors are observed, and the peaks shift to the right to 8. Compared to other particle-resolved studies, our system exhibits a relatively large maximum $Z$ of 14~\cite{dong2022}. It is worth noting that our strongest gel is around 12 times beyond the gel boundary, and the resulting thick and close packed gel arms may contribute to this discrepancy. The polydispersity in our system may also be relevant because larger droplets tend have larger $Z$.

\begin{figure*}[t!]
\centering
\includegraphics[width = 170 mm]{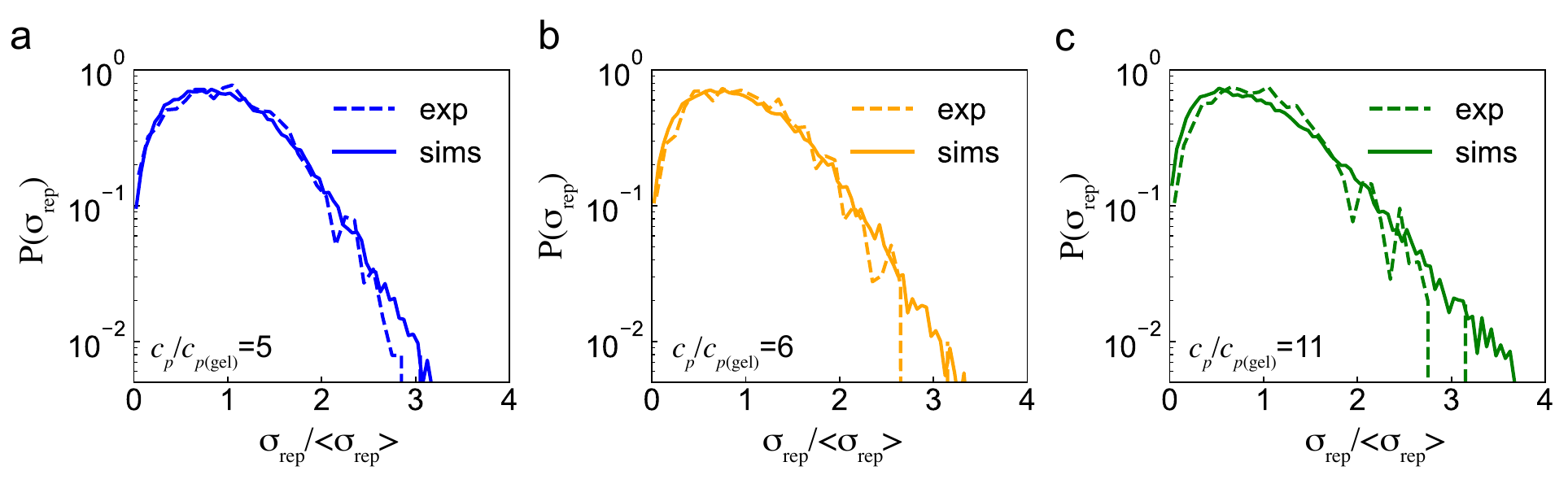}
\caption{\textit{Distributions of repulsive stress.} Experiment and simulation data are compared for the interaction strengths indicated.}
\label{figRepForce}
\end{figure*}

We see similar trends in Fig.~\ref{figNeighboursExp} (b). The state point with $c_\mathrm{p}=0.16$ gl$^{-1}$ display fewer contacts compared to the two higher polymer concentrations. The distribution of the local contact number closely resembles the number of neighbours, indicating that for the gel strength here, most pairs of neighbors have contacts. These contacts provide an estimation of the relative compressive forces acting on droplets. It has been shown that
there are positive correlations between compressive forces and contact sizes \cite{brujic2003}. Therefore we quantify the contact volume $v_\mathrm{c}$ by summing up the number of pixels for each detected contact in 3D, the probability distribution is shown in Fig.~\ref{figNeighboursExp} (c). 
From the Princen model [Eq.~\ref{eqPrincen}],  we expect that interparticle forces $f$ should scale with $v_{c}$ at an exponent of 2/3~\cite{princen1983}. Thus we plot $P(v_\mathrm{c}^{2/3})$ in Fig.~\ref{figNeighboursExp} (d).
We presume that the Princen model holds for our system and we therefore use the distribution in Fig.~\ref{figNeighboursExp} (d) henceforth.}

\subsection{Stress Distribution} 
\label{sectionStressDistribution}

The inferred forces from contacts in Fig.~\ref{figNeighboursExp} (d) are then mapped to the stress using the method described in Sec.~\ref{sectionStress}. This allows us to analyse the local full stress $\sigma_{\mathrm{full}}$, which scales with local pressure, in the simulations as shown in Fig.~\ref{figNeighboursSim} (a).

In Fig.~\ref{figNeighboursSim}(a), we assign the local coarse-grained stress to each particle and plot the resulting distribution. The distributions follow exponential forms, with their widths increasing with polymer concentration. 
We note that we chose to plot the particle-level values; similar distributions were obtained when plotting the coarse-grained stresses calculated over boxes, excluding boxes containing no particles. Similar distributions have previously been reported for gel arms \cite{thijssen2023} and were attributed to residual stresses arising from the dynamically arrested state of colloidal gels.

When measuring the number of contacts in the simulations (Fig.~\ref{figNeighboursSim}(b)), we observe a trend similar to that found in the experiments (Fig.~\ref{figNeighboursExp}): decreasing the interaction strength, which corresponds to a lower polymer concentration, shifts the peak towards smaller contact numbers. However, these shifts are less pronounced in the simulations than in the experiments. This difference may arise from the absence of hydrodynamic interactions in the simulations.

\begin{figure*}[t!]
\centering
\includegraphics[width = 175 mm]{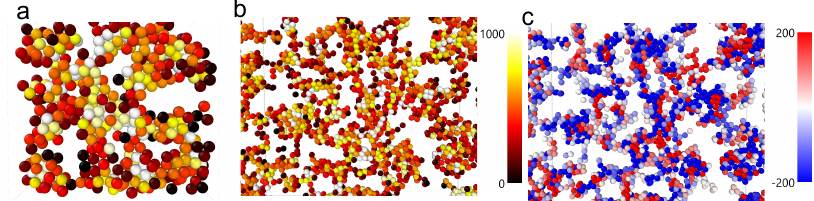}
\caption{\textit{Color coded droplets with stresses.}
(a,b) Renderings of particles \paddyspeaks{in a region of depth} 
two diameters whose colors are rendered as  black, red, and white according to low, medium, and high repulsive stresses for the stress from repulsive forces $\sigma_{\mathrm{rep}}$ $(\mathrm{k_BTd^{-3}})$ for (a) experiments and (b) simulations. The value of the stress is indicated in the colour bar,
(c) Simulation data of particles from the same region as (b). Here the colours are rendered as blue, white, and red according to negative, zero, and positive stress $\sigma_{\mathrm{full}}$ $(\mathrm{k_BTd^{-3}})$. The value of the stress is shown in the colour bar.}
\label{figColourforce}
\end{figure*}

Although the structures differ slightly, the repulsive stress $\sigma_{\mathrm{rep}}$ distributions obtained from experiments and simulations are \rlj{very similar} when the same stress calculation method is applied to both (Fig.~\ref{figRepForce}). The peaks broaden with increasing polymer concentration, possibly because the contacts become larger and more heterogeneous as the droplets are pushed together more strongly. Such peaks have previously been proposed as a signature of yield stress development \cite{ohern2001}, which would be an interesting direction for future investigation.

We again emphasize that the gels experience not only repulsive forces but also an attractive depletion contribution. As noted previously, the attractive contribution is not measured directly in the experiments but can be quantified in the simulations (Fig.~\ref{figColourforce}(c)). We observe a tendency for regions of high repulsive stress to coincide with regions of negative tensile stresses.

The relationship between the measured local full stress $\sigma_{\mathrm{full}}$ and the stress contribution from repulsive forces $\sigma_{\mathrm{rep}}$ in the simulation data is explored further in Fig. \ref{figStressComp}. In particular, the joint probability distribution suggests an association between repulsive-force stress and tensile (negative) stresses, with a higher probability of observing small or negligible pressures. Meanwhile, high repulsive-force stress is frequently observed in regions with compressive (positive) stress. The Pearson correlation coefficient between the two quantities is 0.34, indicating a modest positive linear correlation.

\begin{figure}[t!]
\centering
\includegraphics[width=\columnwidth]{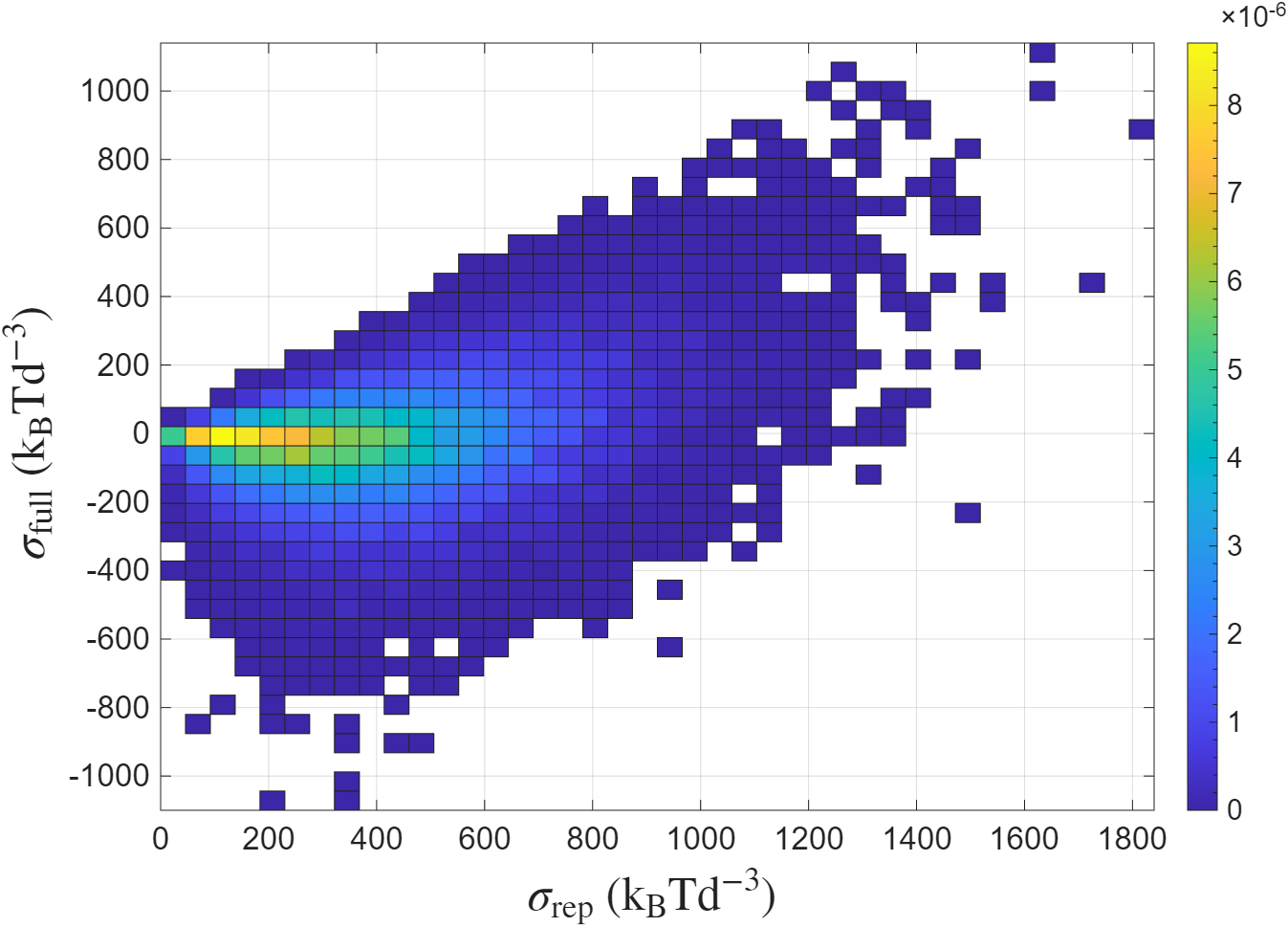}
\caption{\textit{Joint probability distribution of full local stress and stress from repulsive forces.} The joint probability distribution characterises the occurrence of different combinations of full local stress and stress contributions from repulsive forces across all regions of the system. The Pearson correlation coefficient between the two quantities is 0.34, indicating a modest positive linear correlation.}
\label{figStressComp}
\end{figure}

\subsection{The relationship between local structure and stress}
\label{sectionLocal}

The mechanical properties of colloidal gels have been related to minimum energy clusters which are associated with rigidity~\cite{royall2008,royall2021,royall2012}. The topological cluster classification (TCC) was developed to detect structures whose bond network is that of minimum energy clusters~\cite{malins2013tcc}.
The population of selected frequently occurring  TCC clusters in the gels is depicted in Fig.~\ref{figTCCPopulation}, along with the corresponding TCC structures indicated.

\begin{figure*}[t!]
\centering
\includegraphics[width=150mm]{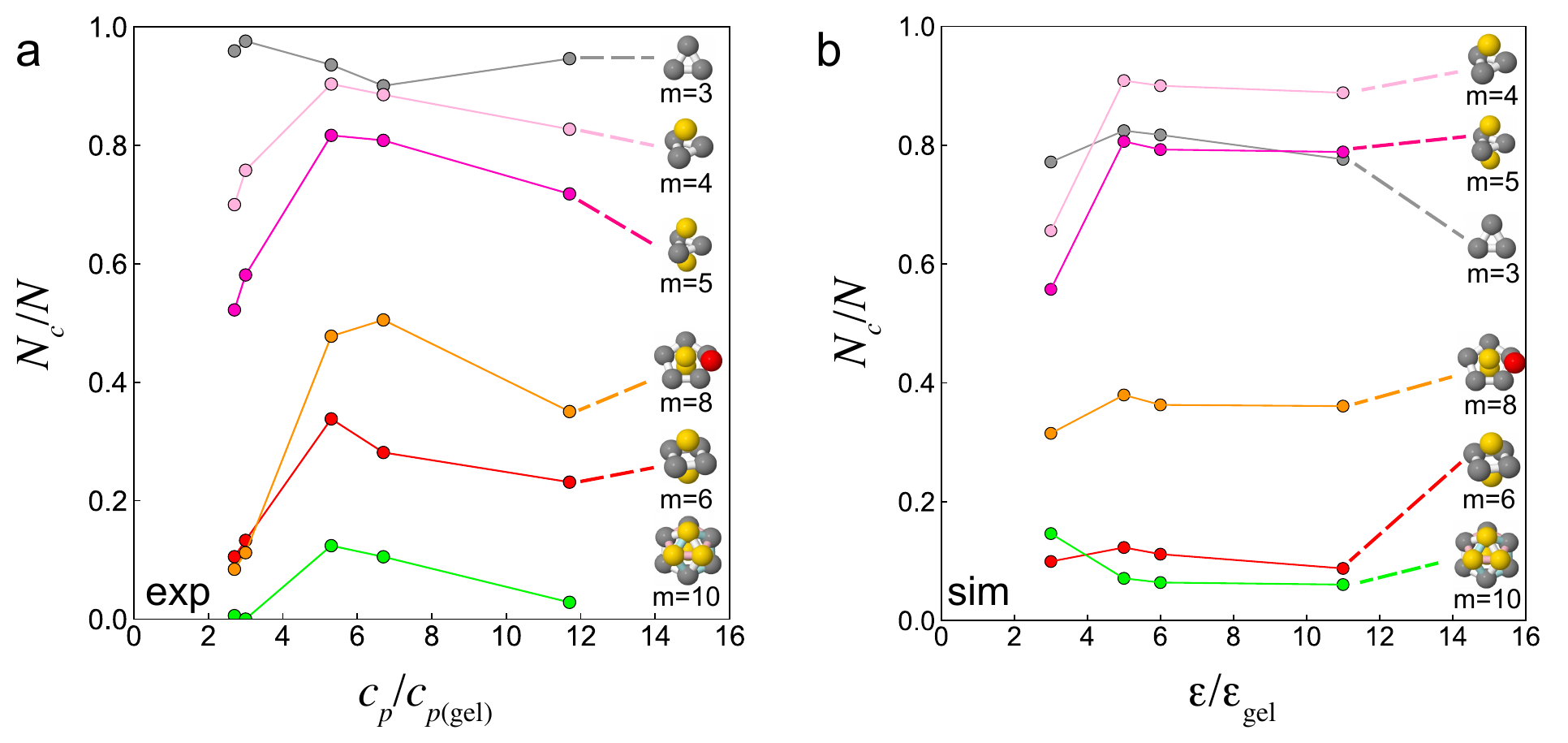}
\caption{\textit{Topological Cluster Classification results.}
Proportion of particles in different TCC structures as a function of polymer concentration, colors denote structures (legend on the right) named by $m$ with the number of particles composing the cluster, followed by its distinguish spatial symmetries for experiments and simulations.}
\label{figTCCPopulation}
\end{figure*}

The dominant local structures are 
$m = 4$  tetrahedra and $m = 5$ triangular bipyramids formed by two tetrahedra. This is followed by medium-sized clusters with $m$ ranging between 6 and 8. Large clusters with $m \gtrsim 8$ are rather rare, as they are suppressed due to rapid dynamical arrest upon gelation in depletion systems~\cite{royall2008}. Such larger clusters like the $m = 10$ defective icosahedra are associated with aging~\cite{royall2018jcp}.

At relatively low polymer concentration $1<c_p/c_{p\mathrm{(gel)}}\lesssim 3$, we noted in Sect.~\ref{sectionStructural} that we saw more Brownian motion and classified these as weak gels. For the TCC clusters in this regime, we see an \emph{increase} in cluster population with the exception of the $m=10$ defective icosahedron in the simulation data.

As the polymer concentration increases from 0.10 gl$^\mathrm{-1}$ to 0.15 gl$^\mathrm{-1}$ ($c_p/c_{p\mathrm{(gel)}}$=3.33 to 5), clear trends show that the population of small clusters decreases and that of large clusters increases. This is consistent with the results from the radial distribution function results in Figure \ref{figG}, the most obvious change indicates the occurrence of gelation. With $c_\mathrm{p}$ increased to 0.20 gl$^\mathrm{-1}$ ($c_p/c_{p\mathrm{(gel)}}$=6.7), we find an increase of basically all cluster populations. Increasing $c_p$ further to 0.35 $g l^\mathrm{-1}$ ($c_p/c_{p\mathrm{(gel)}}$=11.3), only a little descent can be seen in the population of small clusters and a slight increase for that of medium and clusters. This suggests a minor shift from small clusters to medium-sized structures at a deeper quench ($c_p/c_{p\mathrm{(gel)}}$=11.7), which is consistent with previous work which used \emph{solid} sterically stablised polymethyl methacrylate colloids, unlike the emulsion droplets used here~\cite{royall2008,royall2015,royall2018jcp,richard2018}.

\begin{figure}[t!]
\centering
\includegraphics[width = 80 mm]{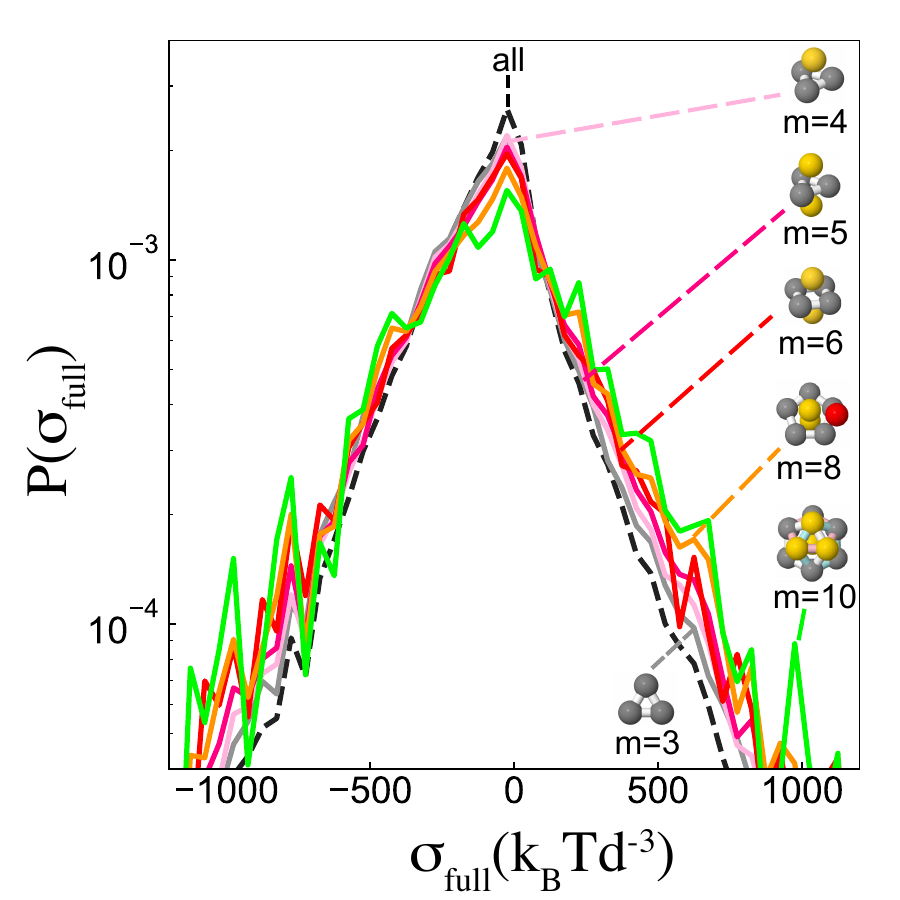}
\caption{\textit{Correlation of stress with rigid minimum energy clusters}.
Here simulation data for the full stress is shown for all particles (dashed line) and particles in certain rigid TCC clusters.}
\label{figFullTCC}
\end{figure}

\begin{figure*}[t!]
\centering
\includegraphics[width = 150 mm]{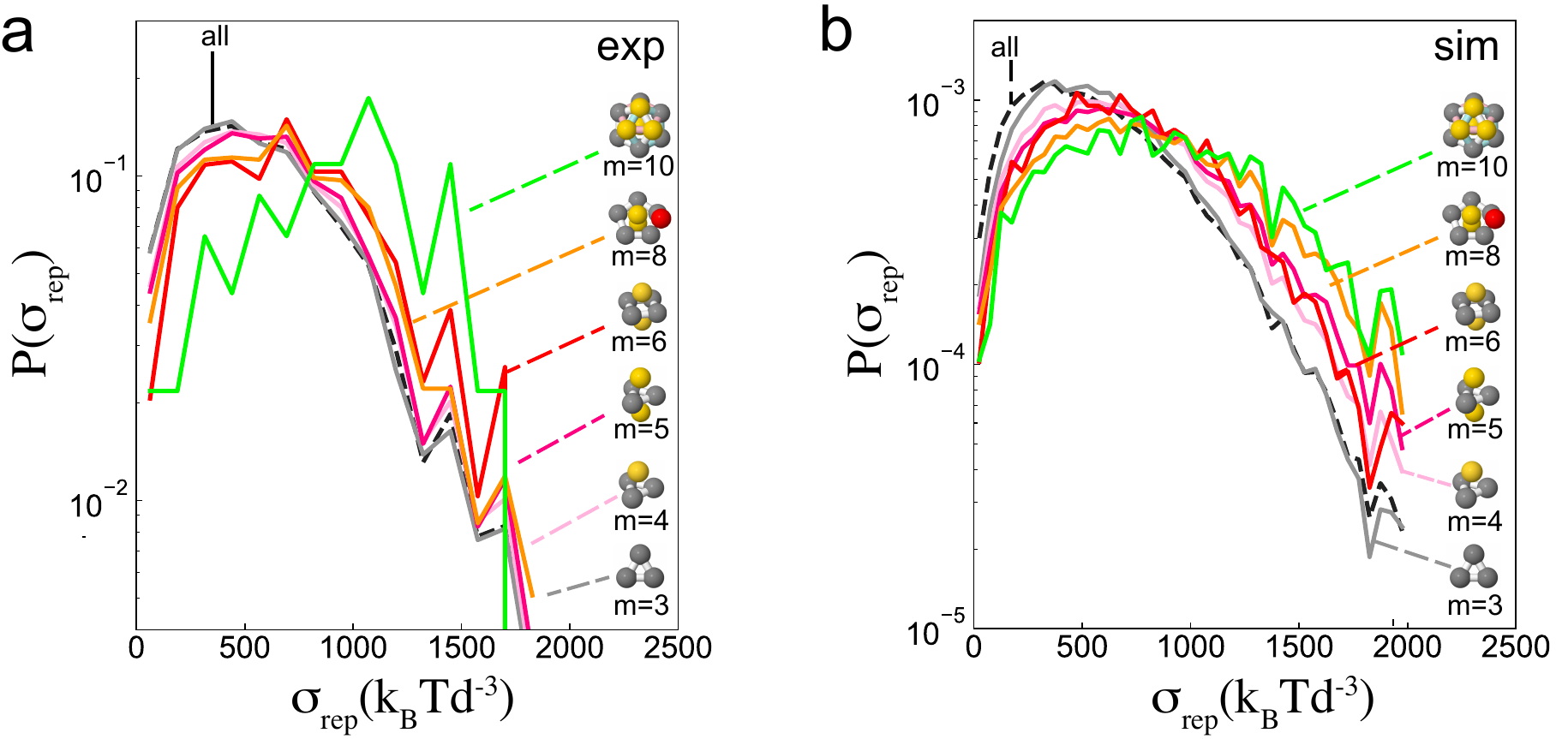}
\caption{
\textit{Distribution of repulsive stress for all particles and for particles in particular rigid local structures.}
(a) Experimental data.
(b) Computer simulation data.
}
\label{figRepulsiveTCC}
\end{figure*}

We now turn our attention to the relationship between local structure and stress. One might expect that the TCC clusters, which are locally rigid, to carry a higher stress than the average. Some evidence for this was found previously~\cite{dong2022}. Here we probe the stress on particles in TCC clusters.  Figure~\ref{figFullTCC}  shows the stress distribution for particles in different TCC clusters.  This is the full stress obtained from simulations, which includes both repulsive and attractive forces. Particles in TCC clusters do indeed have broader distributions, corresponding to local stresses that are larger in magnitude.

In Fig.~\ref{figRepulsiveTCC}, we show similar distributions for the repulsive contribution to the stress, for both experiment and simulation. We see a significant trend, that TCC clusters are associated with higher repulsive stress values. The sightly smaller tail at high stress values in the experiment could be due to limitations in our analysis, for example we have assumed that the Princen model (Eq.~\ref{eqPrincen}) holds. That said, the detailed structure between the experiments and simulations may not be the same, for example due to hydrodynamic interactions~\cite{royall2015,degraaf2019}.

More detailed TCC analyses might include counting the number of tetrahedra in which each particle participates, in order to probe correlations between the stress and the local structure, as in Ref.~\onlinecite{dong2022}.  It would also be possible to examine this correlation with larger TCC structures at longer timescales, or under mechanical perturbations like shear~\cite{bhaumik2025sm}.  For gels at lower colloid volume fraction, load-bearing colloidal strands are more identifiable, which might enable further studies of the correlation between stress and local structure. Time-series data could shed light on how the dynamics of these particles change as stress levels vary, especially during events like particle escape or strand breaking~\cite{thijssen2023,bhaumik2025pre,bhaumik2025sm}.} Another intriguing way to extend this work would be to use solid (eg polymethyl methacrylate) colloids for which \emph{rigidochromic} dyes give similar information to the solvatochromic dye used here~\cite{carpentier}.

\section{Conclusion}
\label{sectionConclusion}

We have presented three key developments in a colloidal model system for measuring stress at the microscopic level. Our method uses STED nanoscopy of an emulsion system to measure droplet coordinates and interdroplet contacts and forces by the use of a solvatochromic dye, nile red. 
First, we have improved the experimental system to allow independent tuning of density and refractive index matching.
Second, we have established a close correspondence between the experiments and computer simulations.
Importantly, this enables us to estimate ensemble-averaged attractive forces between particles, complementing the repulsive component measured experimentally.
Third, by applying the Irving-Kirkwood stress, whose accuracy we have established for our simulations~\cite{thijssen2023}, we are able to analyse local stress in both experiments and simulations within a consistent framework.

We have substantially improved density and refractive index matching in the monodisperse PDMS emulsion system, enabling reliable imaging at greater depths using confocal microscopy. Refractive index matching is optimised by systematically varying the solvent composition, while density matching is independently tuned by adding a dense fluorinated oil to the emulsion droplets. Importantly, these modifications preserve colloidal stability and do not compromise the solvatochromic response of Nile red, which underpins our measurements of interdroplet contacts and forces.

We have established a close correspondence between the experiments and computer simulations, with excellent agreement in both the structural properties and the measured repulsive forces. This agreement supports the use of a relatively simple one-component Brownian dynamics model to capture key features of the experimental system. Previous studies have used similar models to investigate colloidal structure~\cite{royall2008,royall2021,lu2008,royall2015,royall2018jcp,bhaumik2025sm} and bulk mechanical properties~\cite{bonn2017}, whereas our approach extends the comparison to local forces and stresses. Crucially, the simulations provide access to the \emph{attractive} components of the forces and stresses, which cannot be measured directly in our experiments. This allows us to characterise the ensemble-averaged attractive interactions in a model that closely reproduces the experimentally observed structure and repulsive forces.

The accuracy with which the simulations reproduce the experimental results is further demonstrated by the local stress distributions obtained using the Irving–Kirkwood stress formulation. For example, as the attraction strength increases, a shoulder associated with higher repulsive stresses emerges in the experimental distributions and is reproduced in the simulations. Furthermore, the stress distributions resolved by local structural environment reveal a tendency for larger clusters, including those containing around ten particles, to sustain higher stresses. These findings highlight the ability of our approach to capture not only the overall stress distributions but also the relationship between local structure and stress within the gel network.

This work opens the way to further study, in particular this system could be used to analyse local yielding in colloidal gels, for example through the use of a shear stage~\cite{bhaumik2025sm}.

\section*{Acknowledgments}
The authors, especially CPR, would like to warmly thank Hajime Tanaka for his support and great enthusiasm for Soft Matter Physics over many years. Regarding this manuscript, it is a pleasure to thank Daniel Bonn, Roel Dullens, Suzanne Fielding and Matthias Schmidt for stimulating conversations about this work. RC acknowledges support from Bayer AG and the China Scholarship Council. This work was supported by the EPSRC through grants EP/T031247/1 (KT, RLJ) and EP/T031077/1 (AMA, CPR and TBL). CPR was supported by the Agence Nationale de Recherche Grant DiViNew and a Leverhulme Visiting Professorship to the University of Warwick. KT was also supported by Novo Nordisk Foundation (Grant Nos. NNF18SA0035142 and NNF23OC0085012)

\begin{appendix}

\section{Fluorescence Characterization}
\label{sectionFluorescence}

Nile red is a dye with polarity-sensitive characteristics, exhibiting a preference for interfacial regions over hydrophobic areas in water due to its amphiphilic nature. When droplets undergo compression, the spatial confinement within the contact area restrains more dye molecules, leading to even higher intensity. This feature enables a direct correlation between fluorescence and compressive forces \cite{brujic2003,brujic2007}.

Emission spectrum analysis was conducted on PDMS emulsions containing Nile red at different volume fractions \cite{brujic2007}. The local maximum absorption wavelengths were determined for the solely aqueous phase ($\phi = 0$). The peak lies at $\lambda = 640 $ nm, characteristic of the continuous phase. With an increase in droplet volume fraction $\phi$, peaks around $\lambda = 545 $ nm become evident. These shorter-wavelength peaks are attributed to the  droplets, a fact corroborated by the spectra of Nile red in pure PDMS. The intensity of these peaks on the right also grow with increasing droplet volume fraction $\phi$, since the aqueous phase diminishes proportionally, this increase is a result of more interfaces from more droplets.

We carried out similar tests, using a Leica SP8 confocal microscope equipped with a tuneable white light laser. We performed the scan for both droplets 
The corresponding emission wavelengths were 531-565 nm and 597-696 nm, respectively.

We mixed a dense oil 2,3,4,5,6-Pentafluorotoluene (PFT) at a volume ratio of 0.5 to control droplet density, the excitation wavelengths experienced a blue shift to 590 and 630 nm for droplets and interfaces respectively. Correspondingly, the detection wavelengths were moved to 530-580 nm, and 650-710 nm. The intensities of 590 and 630 nm were adjusted to 15 and 100 to reach the desired contact images. 
\end{appendix}

\section*{References}

%

\end{document}